\documentclass[fleqn,12pt,twoside]{article}
\usepackage{espcrc1}
\usepackage{epsfig}
\usepackage{amssymb}

\readRCS
$Id: espcrc1.tex,v 1.2 2004/02/24 11:22:11 spepping Exp $
\usepackage{graphicx}
\usepackage[figuresright]{rotating}

\newcommand{\AmS}{{\protect\the\textfont2
  A\kern-.1667em\lower.5ex\hbox{M}\kern-.125emS}}

\title{Elliptic flow of $\Lambda$ hyperons in Pb+Pb collisions at 158~$A$\hspace{0.08cm}GeV }

\author{Grzegorz Stefanek (for the NA49 Collaboration)}

\begin{document}

\maketitle

\vspace{0.5cm}
\noindent
C.~Alt$^{9}$, T.~Anticic$^{21}$,
B.~Baatar$^{8}$,D.~Barna$^{4}$, J.~Bartke$^{6}$, L.~Betev$^{10}$,
H.~Bia{\l}\-kowska$^{19}$, C.~Blume$^{9}$,  B.~Boimska$^{19}$,
M.~Botje$^{1}$, J.~Bracinik$^{3}$, R.~Bramm$^{9}$,
P.~Bun\v{c}i\'{c}$^{10}$, V.~Cerny$^{3}$, P.~Christakoglou$^{2}$,
O.~Chvala$^{14}$, J.G.~Cramer$^{16}$, P.~Csat\'{o}$^{4}$,
P.~Dinkelaker$^{9}$, V.~Eckardt$^{13}$,
D.~Flierl$^{9}$, Z.~Fodor$^{4}$, P.~Foka$^{7}$, V.~Friese$^{7}$,
J.~G\'{a}l$^{4}$, M.~Ga\'zdzicki$^{9,11}$, V.~Genchev$^{18}$,
G.~Georgopoulos$^{2}$, E.~G{\l}adysz$^{6}$, K.~Grebieszkow$^{20}$,
S.~Hegyi$^{4}$, C.~H\"{o}hne$^{7}$, K.~Kadija$^{21}$,
A.~Karev$^{13}$, M.~Kliemant$^{9}$, S.~Kniege$^{9}$,
V.I.~Kolesnikov$^{8}$, E.~Kornas$^{6}$, R.~Korus$^{11}$,
M.~Kowalski$^{6}$, I.~Kraus$^{7}$, M.~Kreps$^{3}$,
A.~Laszlo$^{4}$, M.~van~Leeuwen$^{1}$, P.~L\'{e}vai$^{4}$,
L.~Litov$^{17}$, B.~Lungwitz$^{9}$, M.~Makariev$^{17}$,
A.I.~Malakhov$^{8}$, M.~Mateev$^{17}$, G.L.~Melkumov$^{8}$,
A.~Mischke$^{1}$, M.~Mitrovski$^{9}$, J.~Moln\'{a}r$^{4}$,
St.~Mr\'owczy\'nski$^{11}$, V.~Nicolic$^{21}$, G.~P\'{a}lla$^{4}$,
A.D.~Panagiotou$^{2}$, D.~Panayotov$^{17}$, A.~Petridis$^{2}$,
M.~Pikna$^{3}$, D.~Prindle$^{16}$, F.~P\"{u}hlhofer$^{12}$,
R.~Renfordt$^{9}$, C.~Roland$^{5}$, G.~Roland$^{5}$, M.
Rybczy\'nski$^{11}$, A.~Rybicki$^{6,10}$, A.~Sandoval$^{7}$,
N.~Schmitz$^{13}$, T.~Schuster$^{9}$, P.~Seyboth$^{13}$,
F.~Sikl\'{e}r$^{4}$, B.~Sitar$^{3}$, E.~Skrzypczak$^{20}$,
G.~Stefanek$^{11}$, R.~Stock$^{9}$, C.~Strabel$^{9}$,
H.~Str\"{o}bele$^{9}$, T.~Susa$^{21}$, I.~Szentp\'{e}tery$^{4}$,
J.~Sziklai$^{4}$, P.~Szymanski$^{10,19}$, V.~Trubnikov$^{19}$,
D.~Varga$^{4,10}$, M.~Vassiliou$^{2}$, G.I.~Veres$^{4,5}$,
G.~Vesztergombi$^{4}$,
D.~Vrani\'{c}$^{7}$, A.~Wetzler$^{9}$, Z.~W{\l}odarczyk$^{11}$
I.K.~Yoo$^{15}$, J.~Zim\'{a}nyi$^{4}$

\vspace{0.5cm} \noindent
$^{1}$NIKHEF, Amsterdam, Netherlands. \\
$^{2}$Department of Physics, University of Athens, Athens, Greece.\\
$^{3}$Comenius University, Bratislava, Slovakia.\\
$^{4}$KFKI Research Institute for Particle and Nuclear Physics, Budapest, Hungary.\\
$^{5}$MIT, Cambridge, USA.\\
$^{6}$Institute of Nuclear Physics, Cracow, Poland.\\
$^{7}$Gesellschaft f\"{u}r Schwerionenforschung (GSI), Darmstadt, Germany.\\
$^{8}$Joint Institute for Nuclear Research, Dubna, Russia.\\
$^{9}$Fachbereich Physik der Universit\"{a}t, Frankfurt, Germany.\\
$^{10}$CERN, Geneva, Switzerland.\\
$^{11}$Institute of Physics \'Swi{\,e}tokrzyska Academy, Kielce, Poland.\\
$^{12}$Fachbereich Physik der Universit\"{a}t, Marburg, Germany.\\
$^{13}$Max-Planck-Institut f\"{u}r Physik, Munich, Germany.\\
$^{14}$Institute of Particle and Nuclear Physics, Charles University, Prague, Czech Republic.\\
$^{15}$Department of Physics, Pusan National University, Pusan, Republic of Korea.\\
$^{16}$Nuclear Physics Laboratory, University of Washington, Seattle, WA, USA.\\
$^{17}$Atomic Physics Department, Sofia University St. Kliment Ohridski, Sofia, Bulgaria.\\
$^{18}$Institute for Nuclear Research and Nuclear Energy, Sofia, Bulgaria.\\
$^{19}$Institute for Nuclear Studies, Warsaw, Poland.\\
$^{20}$Institute for Experimental Physics, University of Warsaw, Warsaw, Poland.\\
$^{21}$Rudjer Boskovic Institute, Zagreb, Croatia.\\

\begin{abstract}
The elliptic flow of $\Lambda$ hyperons has been measured by the
NA49 experiment in semi-central Pb+Pb collisions at
158~$A$\hspace{0.08cm}GeV. The standard method of correlating
particles with an event plane has been used. Measurements of
$v_{2}$ near mid-rapidity are reported as a function of rapidity,
centrality and transverse momentum. Elliptic flow of $\Lambda$
particles increases both with the impact parameter and with the
transverse momentum. It is compared with $v_{2}$ for pions and
protons as well as with models predictions.
\end{abstract}

\section{Introduction}
Elliptic flow at ultrarelativistic energies is interpreted as an
effect of the pressure in the interaction region. It is thus
sensitive both to the equation of state of nuclear matter and to
the degree of thermalization reached in the system. The flow of
heavier particles is particularly interesting as it is less
sensitive to the freeze-out temperature and thus more directly
reflects conditions at the early stage of the collision than the
flow of light particles. In this context the measurement of the
elliptic flow can serve to test various models and gain insight
into the mechanism of the collision at the early stage.

The measurements of the elliptic flow of $\Lambda$ hyperons
supplement earlier results on directed and elliptic flow for
protons and $\pi$ mesons \cite{NA49_PRC68} and represents the
first results on elliptic flow of $\Lambda$ particles at SPS
energies.

\section{Analysis}
The main components of the NA49 detector \cite{NA49_setup} are
four large-volume Time Projection Chambers for tracking and
particle identification by dE/dx measurement with resolution
3$-$6\%. The TPC system deploys two vertex chambers inside the
magnets and two main chambers on both sides of the beam behind the
magnets. Downstream of the TPCs a veto calorimeter detects
projectile spectators and serves as a trigger device and a
centrality selector. The analysed sample consists of 3M
semi-central Pb+Pb events with online trigger selection of the
23.5\% most central collisions. Events were divided into three
from among six previously used centrality bins (table 1 in
\cite{NA49_PRC68}). The measurement in the centrality range
$\sigma/\sigma_{TOT}$ = 5 $-$ 23.5\% is the integral over bins 2
plus 3.

The candidates for $\Lambda$s are selected on a statistical basis
utilizing the kinematic properties of the reconstructed decay
$\Lambda$ $\rightarrow$ $p + \pi^-$ (BR=63.9\%). The
identification method \cite{NA49_Barnby} relies on the invariant
mass cut and daughter track identification by a cut in dE/dx
around the expectation value derived from a Bethe-Bloch
parametrization. The extracted $\Lambda$ candidates have some
background contamination which is below 8\% in our case. The raw
yields of $\Lambda$ hyperons are obtained by counting the number
of entries in the invariant mass peak above estimated background
in every bin of the azimuthal angle with respect to the event
plane. The acceptance of $\Lambda$ hyperons covers the range
$p_{t}$ $\approx 0.4 - 4$ GeV/c and y $\approx$ -1.5 $-$ +1.0. The
elliptic flow analysis used the standard procedure outlined in
\cite{NA49_PRC68,PLB474} to reconstruct the reaction plane for
each event with the necessary corrections for the reaction plane
dispersion. Acceptance corrections are introduced by the
recentering method \cite{NA49_PRC68} and the event mixing
technique with 10 artificial events per one real event.

Presented NA49 results on $\Lambda$ elliptic flow are still
preliminary. Error bars contain statistical errors and uncertainty
of the background subtraction and mixed-events correction.

\section{Results}
The final statistics in Pb+Pb collisions consists of about 1M
$\Lambda$s. It makes possible flow analysis for several rapidity
and $p_t$ bins. Example azimuthal distributions of $\Lambda$
particles with respect to the estimated event plane are shown in
Fig.1 for real and mixed events. Curves represent fits of the
function in the form of a Fourier series with two parameters
$v_{2}$ and $v_{4}$, see eq.(1) in \cite{PLB474}. The
distributions exhibit a strong correlation for real events (solid
symbols, solid lines) and no correlation for mixed-events (open
symbols, dashed lines). The correlation significantly increases
with transverse momentum and also with impact parameter.

\vspace*{-0.8cm}
\begin{figure}[htb]
\begin{center}
\includegraphics[scale=0.45]{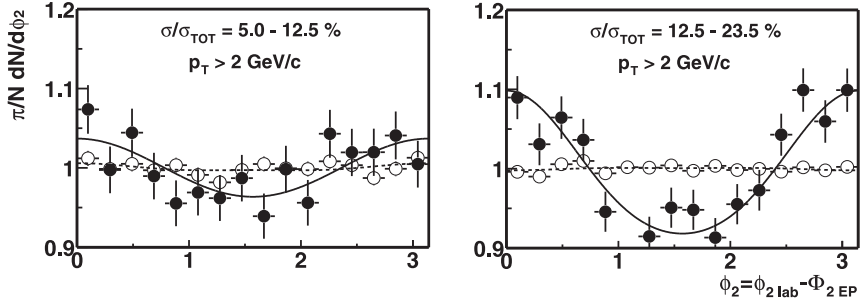}
\vspace*{-0.8cm} \caption{Azimuthal distributions of $\Lambda$
hyperons with respect to the event plane for real events (solid
symbols) and mixed-events (open symbols) in two centrality bins.
The curves are Fourier expansion fits.}
\vspace*{-0.9cm}
\end{center}
\end{figure}
The $p_{t}$ integrated $\Lambda$ elliptic flow exhibits no
significant dependence on rapidity as shown in Fig.2. Similar weak
dependence $v_{2}$(y) for protons in mid-central events is
observed in \cite{NA49_PRC68}(Fig.6).

\vspace*{-0.8cm}
\begin{figure}[htb]
\begin{center}
\includegraphics[scale=0.5]{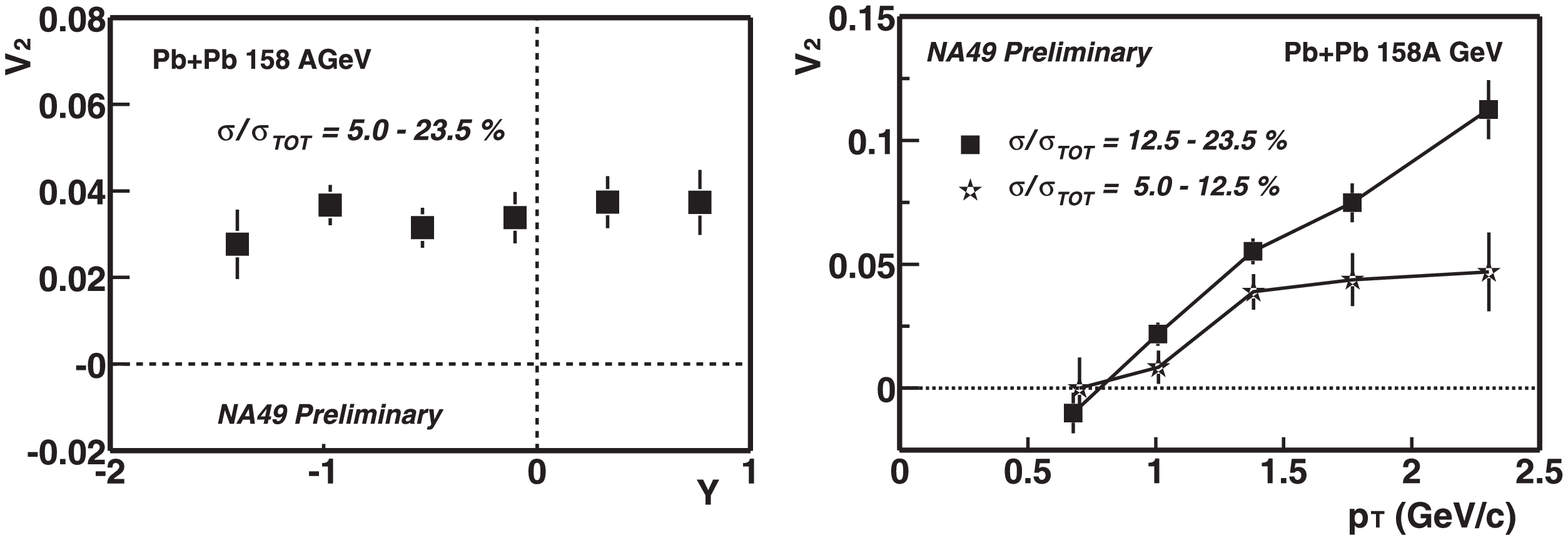}
\vspace*{-0.8cm} \caption{Elliptic flow of $\Lambda$ hyperons as a
function of rapidity (left) and $p_{t}$ (right).} \vspace*{-0.9cm}
\end{center}
\end{figure}
The pronounced flatness of $v_{2}(y)$ suggests that event samples
can be directly compared even in different rapidity ranges as long
as $\Lambda$s are measured near midrapidity. The $p_{t}$
dependence of rapidity integrated $\Lambda$ elliptic flow is shown
in Fig.2(left) for two centrality ranges. The $v_{2}$ parameter
significantly increases with transverse momentum; the rise is
stronger for more peripheral events. The $p_{t}$ dependence of
$\Lambda$ elliptic flow measured by the NA49 experiment is in
agreement with CERES/NA45 data \cite{QM05_CERES}. Fig.3 shows a
comparison of $v_{2}(p_{t})$ of $\Lambda$ hyperons in mid-central
and central events measured by the NA49 and STAR experiments
\cite{lambda_STAR}.

\vspace*{-0.8cm}
\begin{figure}[htb]
\begin{center}
\includegraphics[scale=0.5]{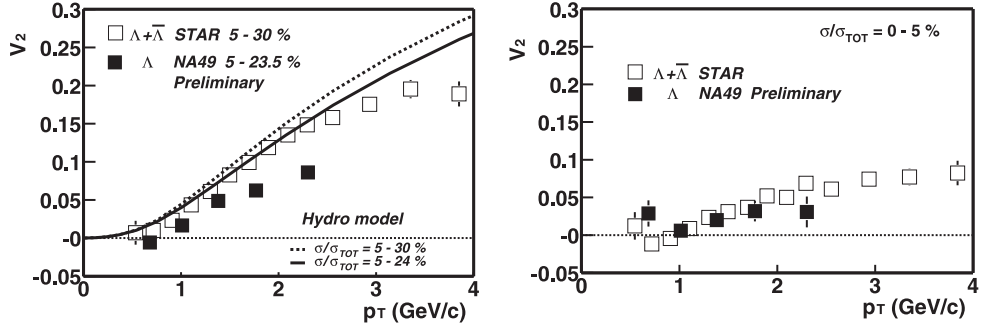}
\vspace*{-0.8cm} \caption{Elliptic flow of $\Lambda$ hyperons as a
function of $p_{t}$ from mid-central (left) and central (right)
events measured by the STAR (open symbols) and NA49 (solid
symbols) experiments. Curves are hydrodynamical model predictions
at RHIC energy.} \vspace*{-0.9cm}
\end{center}
\end{figure}
For mid-central collisions NA49 elliptic flow grows linearly with
$p_{t}$ up to $\sim$2 GeV/c but significantly slower then at RHIC
energy. It has to be emphasized that RHIC mid-central data have
been measured in the centrality range $\sigma/\sigma_{TOT}$ =
5$-$30\% while SPS events are more central. The effect of
different centrality ranges has been estimated by hydrodynamic
calculations \cite{RHIC_Huovinen} at RHIC energy in two centrality
bins shown in Fig.3 as two curves. It only partly explains the
difference between both measurements. For central events both
measurements agree within errors. The elliptic flow $v_{2}(p_{t})$
for pions, protons and $\Lambda$ hyperons measured by the NA49
experiment in mid-central events is displayed in Fig.4.
\vspace*{-0.8cm}
\begin{figure}[htb]
\begin{center}
\includegraphics[scale=0.5]{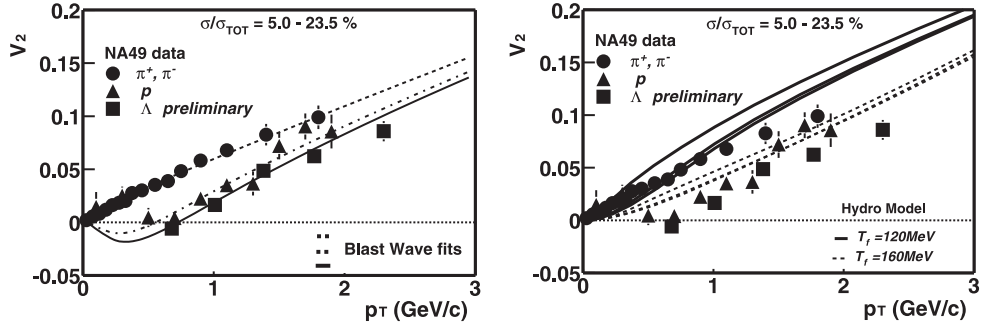}
\vspace*{-0.8cm} \caption{Elliptic flow for charged pions
(circles), protons (triangles) and $\Lambda$ hyperons (squares) as
a function of $p_{t}$ from 158~$A$\hspace{0.08cm}GeV Pb+Pb
mid-central events measured by NA49 experiment. Curves are blast
wave fits (left) and hydrodynamic model predictions for two
freeze-out temperatures at SPS energy (right). } \vspace*{-0.9cm}
\end{center}
\end{figure}
The values for pions and protons are calculated on the basis of
previous results published in \cite{NA49_PRC68}. Curves in
Fig.4(left) indicate fits of the blast wave parametrization
\cite{BW1,BW2} and in Fig.4(right) predictions of a hydrodynamical
model \cite{Huovinen_private}. The elliptic flow grows linearly
with $p_{t}$ for all species but the rise for pions starts from
$p_{t}$ equal zero while for protons and $\Lambda$s it starts from
$p_{t}$=0.5 GeV/c. The elliptic flow for pions is significantly
larger than for heavier particles although at $p_{t}$ $\approx$ 2
GeV/c the magnitude of the flow for all particle species becomes
similar. Data are reproduced by the blast wave fits with
parameters similar to those obtained by fitting pt spectra and HBT
radii \cite{NA49_HBT}. Hydrodynamical model calculations assume a
first order phase transition to QGP at critical temperature
$T_{c}$=165 MeV.  The freeze-out temperature $T_{f}$=120 MeV is
tuned to reproduce particle spectra. Model calculations
significantly overestimate the SPS data for semi-central
collisions in contradiction to predictions at RHIC energy which
agree with data quite well for $p_{t}$ $\lesssim$ 2GeV
\cite{lambda_STAR}. The discrepancy at SPS may indicate a lack of
complete termalisation or a viscosity effect. On the other hand
the model reproduces the characteristic hadron mass ordering of
elliptic flow and thus supports the hypothesis of early
development of collectivity. The calculation from the same model
with higher temperature $T_{f}$=160 MeV exhibits better agreement
with $\Lambda$ flow data. This might be considered as an
interesting check of the early decoupling scenario for hyperons,
but the model with such a high freeze-out temperature has a
problem to simultaneously reproduce $m_{T}$ spectra and $v_{2}$
values.

\section{Summary}
The NA49 collaboration has measured $\Lambda$ hyperon elliptic
flow at the highest SPS energy. Elliptic flow of $\Lambda$
hyperons exhibits no significant dependence on rapidity for
$y=y_{mid}\pm$1.5. It rises linearly with $p_{t}$ and is smaller
than $v_{2}$ for pions and protons. Both features are well
reproduced by the blast wave parametrization and the hydrodynamic
model. The increase of $v_{2}$ with $p_{t}$ is weaker at SPS than
at RHIC energy and is significantly overpredicted by
hydrodynamical calculations.

\end{document}